\documentclass[twocolumn]{aa} 
\usepackage{graphicx}
\input{epsf}
\input epsfig.sty
\usepackage{txfonts}
\begin{document}
   \title{Was the Narrow Line Seyfert 1 RGB J0044+193 ever radio loud?}

   \author{T.J. Maccarone\inst{1} \and J.C.A. Miller-Jones\inst{1}
          \and
          R.P. Fender\inst{2,1}
	  \and
	  G.G. Pooley\inst{3}
          }

   \offprints{T.J. Maccarone}

   \institute{Astronomical Institute ``Anton Pannekoek,'' University
              of Amsterdam, Kruislaan 403, Amsterdam, The Netherlands,
              1098 SJ\\
              \email{tjm@science.uva.nl,jmiller@science.uva.nl} \and
              Department of Physics and Astronomy, University of
              Southampton, Hampshire SO17 1BJ,United Kingdom\\
              \email{rpf@phys.soton.ac.uk} \and Mullard Radio
              Astronomy Observatory, Cavendish Laboratory, Madingley
              Road, Cambridge CB3 0HE, UK\\ }

   \date{}

   \abstract{We show new radio data and a re-analysis of old data for
   the Narrow Line Seyfert 1 (NLSy1) galaxy RGB~J0044+193.  This
   galaxy has previously been suggested to be both radio loud, and
   highly variable in the radio.  As most NLSy 1 galaxies are radio
   quiet, this was interpreted as possible evidence that this AGN's
   jet was beamed toward the observer.  This object was probably
   never either radio loud and nor variable -- the one exception to
   this rule is a marginal detection of the object in the 1987 Green
   Bank survey, which we argue may have been spurious.}

   \maketitle
%

\section{Introduction}

In recent years, a growing understanding of the coupling between
accretion disks and relativistic jets has been developing.  In
particular, it has been found that the ratio of monochromatic radio
flux density to accretion luminosity increases with black hole mass
and decreases with increasing luminosity in Eddington units (see
e.g. Boroson 2002; Gallo, Fender \& Pooley, 2003; Merloni, Heinz \& Di
Matteo 2003 -- MHDM03; Falcke, K\"ording \& Markoff 2004 -- FKM04).
In X-ray binaries, it has become well established also that the radio
emission is suppressed when the accretion disk enters into a high/soft
state (e.g. Tananbaum et al. 1972; Fender et al. 1999) -- i.e. when
the X-ray spectrum is well fit by a geometrically thin disk model
(e.g. Shakura \& Sunyaev 1973) rather than by a hard power law
component.  More recently, this has also been shown to be likely in
AGN with soft X-ray spectra and luminosities of a few percent of the
Eddington limit (Maccarone, Gallo \& Fender 2003 -- MGF03).
Theoretical models also predict suppression of the radio emission in
the soft state -- large scale height magnetic fields are typically
needed to extract power from either the accretion disk or the black
hole's spin into a relativistic jet (e.g. Livio, Ogilvie \& Pringle
1999; Meier 2001).

On the other hand, the velocity of a matter jet (i.e. one not
dominated by Poynting flux) should be approximately the escape
velocity of matter from the orbital radius where the jet is ejected.
The lowest luminosity states, in which the geometrically thick part of
the accretion disk is thought to extend the furthest away from the
compact object (e.g. Esin, McClintock \& Narayan 1997), should thus
have the lowest jet velocities (e.g. Meier 1999).  Observations of the
evolution of jet properties in Galactic X-ray transients seem to
indicate at least that the jet velocities seen when sources are at
high X-ray luminosities are higher than the jet velocities seen at low
X-ray luminosities (Fender, Belloni \& Gallo 2004).  It is thus
possible, and perhaps quite reasonable, that a substantial fraction of
the jet suppression in high/soft state systems is not because the jet
is intrinsically weaker, but rather because the jet velocity is quite
high and the jet will be Doppler de-boosted, except on lines of sight
very close to our own.  Such jets, when pointed at us, might be
expected to have rather remarkable properties, and to allow us to
probe new physics regimes not allowed by more normal jets, so searches
for them are quite important.

One such source (RGB~J0044+193) had been argued to show evidence for
strong variability, as it was seen at a level of 24 mJy at 5 GHz in an
update to the 1987 Green Bank (87GB) survey (see Gregory \& Condon
1991 for discussion of the survey; see also the updates to the catalog
of Neumann et al. 1994; Brinkmann et al. 1995), undetected at 1.4 GHz
in the 1992 NRAO VLA Sky Survey (NVSS -- see Condon et al. 1998)
observations (which implied an upper limit of about 2.5 mJy), and
detected again in a Very Large Array (VLA) follow-up survey of the
1987 Green Bank sources, with a flux density level of 7 mJy at 5 GHz
(Laurent-Muehleisen et al. 1997).  We direct the reader to Siebert et
al. (1999) for a summary of the past radio observations and arguments
that these observations imply a variable source.  In this research
note, we will present a re-analysis of the VLA follow-up observations,
showing a source at a flux density level of about 0.8 mJy, as well as
data from new Westerbork observations showing the same source at a
flux density level of about 0.7 mJy.  We also show that there is
marginal evidence for the source in the NVSS data (albeit at a flux
density level below the threshold for listing in the NVSS catalogs),
and that this flux density level is consistent with our detection of
the source at 2.0 mJy at 1.4 GHz in new Westerbork data, and note that
the source was undetected in several observations with the Ryle
telescope at 15 GHz, consistent with the extrapolation of the radio
spectrum obtained with Westerbork and the hypothesis of little or no
radio variability.  We thus consider the possibility that the 87GB
detection was spurious and that this object is not strongly variable
and has never been radio loud.

\section{Data used, analysis procedure, and results}
The dates, frequencies and flux densities of the observations used in
this paper are shown in Table 1.  Below appears a description of the
reduction procedures.  We note that RGB~J0044+193 and the nearby
background sources appear to be point-like within the instrumental
angular resolutions, so we assume the flux density in mJy to be equal
to the peak flux density in mJy/beam for each source.

\begin{table}
\begin{tabular}{lccccl}
\hline
Obs. &Freq.&Date&Flux&Notes\cr
     &     &    &dens.&\cr
\hline
GB & 5 GHz& 1987 & 24 mJy& (1)\cr
VLA& 5 GHz&1994 Sep 16& 0.8 mJy & (2)\cr
WSRT& 5 GHz&2003 Nov 29-30 & 0.8 mJy & \cr
WSRT& 1.4 GHz&2003 Nov 29-30 & 2.0 mJy & \cr
VLA& 1.4 GHz&1991 Feb 1 &non-detection& (3)\cr
Ryle& 15 GHz& 2003 &0.31 mJy \cr
\hline
\end{tabular}
\caption{The radio data used in this paper -- the observatory used, the
frequency at which the data was taken, the date of the observations,
the measured flux density, and a comment flag.  Comments: (1) The
Green Bank point may be spurious for reasons discussed in the text.
(2) This flux density was reported to be 7 mJy, but we argue in the
text that this report was erroneous, and (3) this is the data point
from the NVSS; the images show a possible source at 1.5$\pm$0.5 mJy.}
\end{table}

\subsection{Westerbork Observatory}
We observed RGB~J0044+193 with the Westerbork Radio Synthesis
Telescope for 24 hours on source (12 hours at 1.4 GHz and 12 hours at
5 GHz), plus two hours of calibration time, on November 29 \& 30,
2003.  The data was analyzed using the MIRIAD package (Sault, Teuben
\& Wright 1995).  The source was detected at both frequencies, with a
flux density level of 2.0 mJy at 1.4 GHz, and 0.8 mJy at 5 GHz.  The
noise levels were 10 $\mu$Jy at 1.4 GHz and 70 $\mu$Jy at 5 GHz.
Allowing for some mild possible errors in absolute flux density
calibration, we estimate that the errors should be no more than about
0.1 mJy in both bands.  The implied spectral index from these two data
points is $\alpha=-0.7$, where the flux density, $f_\nu$ is expressed
as a power law function of the frequency such that $f_\nu\propto
$$\nu^\alpha$. In Figure 1, we present the 5 GHz Westerbork image.

\subsection{VLA}

The VLA has made three observations of fields including
RGB~J0044+193 -- two at 5 GHz and one at 1.4 GHz.  One of the 5 GHz
observations was quite short (90 seconds) and contained no suitable
phase calibrator (the closest phase calibrator on the sky was 45
degrees away), so we have not analyzed it.  The observation at 1.4 GHz
is the NVSS observation; no source appears in the NVSS catalog (Condon
et al. 1998).  The final observation was from the survey of
Laurent-Muehleisen et al. (1997), which claimed a detection of this
object at 7 mJy.  We have reanalyzed this observation, calibrating and
imaging the data using standard procedures and the NRAO AIPS data
reduction package.  3C\,286 was used as the primary calibrator, and
J0042+233 (4$^{\circ}$ from the target) as the secondary
calibrator.  Imaging was carried out using natural weighting in order
to maximise the signal-to-noise ratio.  One round of phase-only
self-calibration was carried out before making the final image.  It
was necessary to image a large field ($13^{\prime}\times
13^{\prime}$), since it was found that RGB J\,0045+193 lay on the
sidelobe of a 1.2-mJy source $5.5^{\prime}$ to the northeast of
the target, and it was important to be able to accurately remove the
sidelobes in order to measure the target source flux density.

We detect RGB~J0044+193 at a flux density level of 765$\pm$91$\mu$Jy.
It is likely that the report of a 7 mJy flux density in previous work
was the result of a typographical or other clerical error
(S. Laurent-Muehleisen, private communication).  Two other sources in
the field appear with consistent flux densities in the both of 5-GHz
observations with the WSRT and our re-reduction of the VLA data,
providing confidence in the measurements of J0044+193.

\begin{figure}
\epsfig{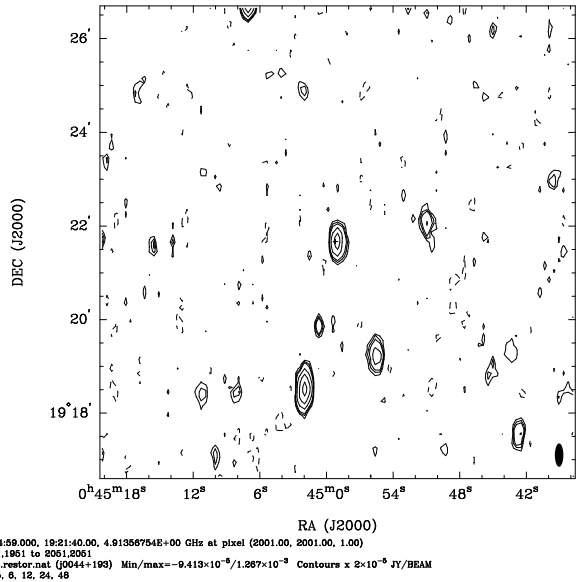}
\caption{Left: The Westerbork observations at 5 GHz.  The VLA map
looks essentially the same, except that it has higher angular
resolution.}
\end{figure}

\subsection{Ryle Telescope}
The Ryle Telescope observed RGB~J0044+193 at 15 GHz 24 times from May
through August of 2003 (see Pooley \& Fender 1997 for a description of
the data reduction procedures).  No detections were made.  The
averaged flux density at the source position is about 0.31$\pm$0.15
mJy, consistent with the $\alpha=-0.7$ power law spectrum measured
with Westerbork.

\section{Comparison with past results and discussion}
Visual inspection of the publicly available NVSS FITS images shows a
1.5 mJy source at 1.4 GHz at exactly the position of RGB~J0044+193.
While this is not a convincing detection taken on its own (it is about
3$\sigma$ above the noise), it is within 1$\sigma$ of the flux density
of the source at 1.4 GHz found with Westerbork, indicating that large
scale variability at 1.4 GHz was unlikely over the 11-year period
between the NVSS data point and the Westerbork observation.  The flux
densities measured by the VLA in 1994 and by Westerbork in 2003 at 5
GHz are also consistent with one another at the 1$\sigma$ level, and
the 3$\sigma$ upper limit on the variability level is only about 40\%.
It seems rather unlikely that source would vary by a factor of 30
between 1987 and 1994, while showing no evidence for variability
between 1994 and 2003.  On the other hand, little is known about jets
from soft state accretion disks, so these systems might exhibit brief
transient episodes.

Apart from this evidence for constancy of the source over an 11-year
baseline, there is also some evidence that the 1987 Green Bank survey
detection of RGB~J0044+193 was spurious.  First, we note that this
object was not in either the original 87GB catalog (Gregory \& Condon
1991) or the follow-up of that catalog using a second epoch of data
(Gregory et al. 1996).  The nearest catalogued source to the position
of RGB J0044+193 is a 34 mJy source about 11 arcminutes away, with a
positional uncertainty of about 20 arcseconds, and the sensitivity
limit reported by Gregory et al. (1996) is about 22 mJy at a
declination of +19 degrees.  The Green Bank catalog was also examined
by Neumann et al. (1994) who produced an additional catalog of sources
in that data, but RGB~J0044+193 did not appear in their published
catalog; the first publication of the radio flux density of
RGB~J0044+193 appeared in Brinkmann et al. (1995), who found a flux
density value of 24 mJy with an error in the 14-23\% range.  We have
communicated directly with W. Brinkmann and he is confident that the
source detection was real, but is unable to recall so long after the
fact why it appeared in only one of these two papers.  However, this
detection corresponds to a 4 to 7 $\sigma$ detection level.  Since the
87GB survey covered 20,000 square degrees with about 2 arcminute
angular resolution, some spurious 4$\sigma$ detections are likely; the
exact amount is difficult to quantify because the sensitivity limit is
due to source confusion rather than Gaussian noise.

The basis for the previous claims that RGB~J0044+193 was a radio loud
AGN was the flux density ratio of 31 between 5 GHz and the optical
magnitude (Siebert et al. 1999), where the cutoff between radio quiet
and radio loud is defined to be a ratio of 10 (Kellermann et
al. 1989). The reduction of the maximum radio flux density by a factor
of 30 implied by these results yields a ratio of radio-to-optical flux
of about 1.  We are thus left with two possibilities -- that the
source was flaring when it was observed during the 1987 Green Bank
survey, and hence that its radio variability is among the very
strongest for all AGN, or that the detection in the 1987 Green Bank
data was spurious, and hence that this source was never radio loud and
shows no signs of radio variability.  While we consider the latter
more likely, we believe that the former cannot be excluded at the
present time.  Future radio monitoring of this source would be quite
useful; we also note that this source should be brighter than
predicted sensitivity limits of the Low Frequency Array (R\"ottgering
2003), so in the near future, detection of flaring activity would not
require a dedicated monitoring program.

\begin{acknowledgements}
We are grateful to Elena Gallo and Dave Meier for useful discussions
about jets from soft state systems; Sally Laurent-Muehleisen for
useful discussions about the VLA follow-up survey of the 1987 Green
Bank sources and for providing data to us, Margo Aller for providing
data about the most variable AGN in the University of Michigan
monitoring source lists, and to the Westerbork Radio Observatory staff
for flexible scheduling.  This work uses archival data from the VLA,
part of NRAO.  The NRAO is a facility of the National Science
Foundation operated under cooperative agreement by Associated
Universities, Inc.  The Westerbork Synthesis Radio Telescope is
operated by the ASTRON (Netherlands Foundation for Research in
Astronomy) with support from the Netherlands Foundation for Scientific
Research (NWO).  The Ryle Telescope is supported by PPARC.
\end{acknowledgements}

\end{document}